\documentclass{article}
\begin{document}
\title{\Huge Fundamental units of length and time}
\author{A. N. Bernal, M. P. L\'opez and M. S\'anchez\footnote{Part of the results of this article has been announced at the meeting ``XXIV International Workshop on the Fundamental Problems of High Energy Physics and Field Theory'' 27-29 June 2001, Protvino, Russia.}}
\date{}
\maketitle
\vspace*{-4mm}
\begin{center}
{\small Dpto. de Geometr\'{\i}a y Topolog\'{\i}a, Universidad de Granada, \\Facultad de Ciencias, Fuentenueva s/n, E--18071 Granada, Spain. \\Email: sanchezm@goliat.ugr.es}
\end{center}
\vspace*{4mm}
 
\bigskip
\begin{quote}
{\normalsize 
Ideal rods and clocks are defined as an infinitesimal symmetry of the spacetime, at least in the non-quantum case. Since no {\rm a priori} geometric structure is considered, all the possible models of spacetime are obtained.} 
\end{quote}

\section{Introduction}
In order to determine how space and time are measured, there are two standard approaches: (A) to postulate the existence of ideal measure instruments (ideal clocks and ideal rods or yardsticks, or some type of rigid bodies) and (B) to assume the existence of priviledged physical particles (ideal freely falling particles, light rays). This second approach is followed by Ehlers, Pirani and Schild (EPS in what follows, see Ref. \ref{EPS} or Ref. \ref{Eh}). It seems to be more natural from a physical viewpoint; in fact, EPS has become a standard foundation for General Relativity. However, one can find some limitations of EPS (or any previous approach based in either (A) or (B)):

\begin{enumerate}
\item  The list of EPS axioms is long. Essentially these axioms: (a) take
events, light rays and freely falling particles as primitive concepts, Ref. \ref{Eh}, Sec. 2.2, (b) postulate radar coordinate systems; in particular, one has  charts, differentiable atlases and a manifold structure, Ref. \ref{Eh}, Sec. 2.3, axiom $L_1$, and (c) ensure a good behaviour for a conformal structure, a projective structure and the compatibility between them, Ref. \ref{Eh}, pp. 26---31; axioms $L_2, L_3 , F_1, F_2, C$, in addition to a chronological order, Ref. \ref{Eh}, axiom $P$. However, these axioms are not enough to ensure the compatibility of these structures with a metric and, therefore, this compatibility is assumed additionally, Ref. \ref{Eh}, formula (2.21). Thus, this approach seems too close to the geometrical model (the Lorentzian metric) that will be obtained finally.

At any case, recall that light clocks can be constructed from these axioms, and it is possible to proceed without rigid bodies (ideal rods or yardsticks can be understood as approximations in tangent space). If one does not follow EPS approach, the alternative approach (A) postulates directly the existence of ideal clocks and yardsticks, and this seems more complicated (or less physically reasonable). 

\item EPS axioms are introduced for General Relativity. But it should be nice to have a set of postulates with no assumed physical theory {\em a priori}, that is, applicable not only to General Relativity but also to Newton theory, and even to any physically reasonable theory for space and time. This also holds for ideal instruments: in principle, they are postulated for a concrete (Newton, Einstein) theory of spacetime. 

\item In General Relativity and, thus, in EPS, one can wonder about the priviledged (but asymmetric) role played by gravity and electromagnetism on the structure of spacetime. Of course, this does not happen in Newton's theory (spacetime is a {\em scenario} for all the forces, including gravity and electromagnetism). In alternative theories as the one by Logunov, Ref. \ref{Lo}, electromagnetism has a priviledged role, but not gravity. Thus, one can wonder how many forces must be included as properties of spacetime.

Moreover, one can wonder if, beyond the ``primitive concepts'' (freely falling particles, light rays) which live in spacetime, there is not a more elemental step concerning the possibility of measuring spacetime itself.  

\end{enumerate}
In this article, we face these problems and the dichotomy (A)-(B) from a simpler viewpoint. We assume a basic fact (Postulate P2) which is independent of any  theory of spacetime (at least, under a non-quantum approach): P2 {\em expresses just the {\em possibility} that, at any case, standard instruments can be constructed, for any theory of spacetime.} From this assumption and two natural postulates
more (``ambient'' assumption P1, differentiability P3) we deduce all possible general models of spacetime. We obtain, essentially,  only four possible models, or even just the following unified one. Spacetime must be a 4-manifold endowed with a metric, perhaps signature--changing or degenerate. This metric is Lorentzian in some open subset, Riemannian in another, may degenerate in tangent space (in this case, with an {\em anti--Leibnizian} structure) and may degenerate in cotangent space (in this case, with a {\em Leibnizian} structure, which generalizes classical Newtonian structures).

Our viewpoint can be summarized as follows. Each observer $O$ takes a coordinate system on spacetime $(U,\Phi)= (t,x^1,x^2,x^3)$. In principle, it is not relevant the exact method used by the observer  for obtaining those coordinates. Elastic rods and non-periodic clocks could be used as proper instruments for measuring the length and time intervals. Nevertheless, if there is a second observer $\tilde O$ with $(\tilde U, \tilde \Phi)= (\tilde t, \tilde x^1, \tilde x^2, \tilde x^3)$, their instruments at each point can be compared between them as follows: {\em observers $O$ and ${\tilde O}$ measure at $p$ using the same unit of time, if the ${\tilde O}$ time, measured using the $O$ clock, goes by as the $O$ time, observed by the ${\tilde O}$ clock (i.e.: if $\partial_{\,t}\,{\tilde t}|_p=\partial_{\,{\tilde t}}\,t|_p $)}. Similar reasons will be argued for units of length, using spatial coordinates, i.e.: {\em $O$ and ${\tilde O}$, measure at $p$ using the same unit of length, if $\partial_{\,x^j}{\tilde x}^i|_p=\partial_{\,{\tilde x}^i}x^j|_p,\,\;\forall i,j\in\{1,2,3\}$}. This suggests that two observers are \emph{compatible with a fundamental system of units at $p$} (or simply, \emph{compatible}), if and only if, its temporal and spatial crossed derivatives coincide. This definition of fundamental units is the central point of this work (Section {\rm 2}). 

As a direct consequence, we obtain (Section {\rm 3} and Section {\rm 4}) the following possible models of spacetime:

\begin{itemize}
\item {\em Temporal Models: } A structure depending on a parameter $c\in [0,\infty]$, the latter being interpreted as the supremum of relative speeds between observers, usually called the \emph{speed of light}. This structure is:
\begin{enumerate} 
\item A Lorentzian scalar product when $c\in (0,\infty)$. So, if this structure is maintained at every point $p$, then a Lorentz metric on all the manifold is obtained.
\item A non-null linear form (up to a sign) and a Euclidean scalar product on its kernel, when $c=\infty$. Therefore, if this occurs at every point $c\equiv\infty$, then a \emph{Leibnizian structure} is obtained. This structure (an everywhere non-zero  $1$-form $\Omega$ and a Riemannian metric in its kernel) is a generalization of the Newtonian classic: notice that the curvature of the Riemannian bundle is arbitrary, and only when $\Omega$ is exact ($\Omega = dt$), terms such as ``synchronizability'' or ``universal time'' will be  employed without any trouble.
\item A non-null vector (up to a sign) and a Euclidean scalar product on its kernel in dual space, when $c=0$. So, an {\em anti-Leibnizian structure} is obtained if $c\equiv 0$. This case is mathematically analogous to the previous one, although physically it is completely different: all the standard observers are at relative rest, however, they can measure different temporal coordinates.
\end{enumerate}

\item {\em Riemannian Model: } A Euclidean product, also depending on a parameter $k>0$. Thus, if this holds at every point, then a Riemannian metric (and a positive function) is obtained over all the manifold.

\item {\em Residual Model: }
A residual case consisting essentially of a set of at most four Euclidean products. If this holds at every point, then a set of at most four  Riemannian metrics (which may coincide in some subsets), depending on positive parameters, is obtained.
\end{itemize}

In the frame of the temporal models, there is nothing impeding $c$ (and the corresponding mathematical structures) a smooth variation from a point to another one within $[0,\infty]$. Moreover, temporal, Riemannian (or even temporal, Riemannian and residual) models can co-exist. Essentially, putting $k=-c^2$ in the temporal models, $k$ varies smoothly in the circle $\mathbf{R} \cup\{\omega\}$ ($\omega=\pm\infty$), obtaining a signature-changing metric over all the manifold (Lorentzian in an open subset, Riemannian in another open subset, being both regions separated by  connected closed subsets, with $k\in\{0, \omega \}$  constant in each one). Therefore, a manifold endowed with such a signature-changing metric (and perhaps with some splittings in the Riemannian part to cover the residual case) is the more general possible model of spacetime. Additionally, the Leibnizian or anti-Leibnizian structures on the degenerate parts as well as function $k$ over all the manifold (with possible splittings in the Riemannian parts) are obtained.

Finally, it should be also pointed out that the parameter $k$ above is either determined unequivocally over the manifold, or can be arbitrarily chosen at all points. In the latter, a local $1+3$ product structure is obtained, which is compatible with all temporal or Riemannian models. This degenerated structure is a space and time classical model for which all observers have an absolute velocity.

This article is organised as follows. The postulates of our theory are introduced and discussed in Section 2. From these postulates, a manifold structure $M$ is obtained and, for each event $p\in M$, a set $S_p$ of preferred basis in the tangent space $T_pM$ can be chosen. In Section 3 we characterize the structure of each $S_p$ (Theorem 3.1). In Section 4 we study the mathematical structures compatible with the structure of $S_p$, and deduce the former models of spacetime. Further discussions of our postulates are given in Section 5, where we discuss possible additional postulates. Finally, our conclusions are summarized in Section 6.

\section{Postulates of the theory}
In this section, postulates defining the fundamental units of time and length are listed. As mentioned above, they are based on the spacetime infinitesimal structure.

\subsection{First postulate, P1}
Our starting point is the following. Spacetime is a set of points or ``events''
with no geometric structure {\rm a priori}. But spacetime can be ``observed''
and the observers parametrize events in a neighborhood around a point. The specific procedure used to obtain this parametrization will not be relevant, even though we will point out  that the time coordinate is distinguishable of the space ones. At any case, the local possibility to parametrize events determines a manifold structure for spacetime.\\

\noindent{\bf P1 (Spacetime and observers) }{\em Spacetime} $M$ is a set of points, called {\em events}, endowed with a structure of differentiable 4--manifold. For any $p\in M$, any coordinate system $O\equiv (U,\Phi)=(t,x^1,x^2,x^3)$ centered at $p$ (i.e.: $\Phi(p)=(0,0,0,0)$) will be called {\em observer around $p$}; the first coordinate $t$ is called {\em the temporal coordinate} of $O$ and the other three $(x^1, x^2, x^3)$ the {\em spatial coordinates}.\\

\noindent{\em Discussion {\rm 1}}. (i) Strictly, this postulate only says that spacetime is a 4-manifold. At this level, the other elements are just definitions (event, observer, temporal and spatial coordinate). They are written just to make easier the mathematical abstraction. (ii) From a physical viewpoint, perhaps the name of ``observer'' for any coordinate system (chart) may sound somewhat strange. On one hand, it may seem too general: one can think that just some of the charts may correspond to real physical observers. Nevertheless, this is not an obstacle to admit P1 in principle. The relevant restrictions on the observers will be introduced in the second postulate. On the other hand, the word ``observer'' may have unappropriate connotations for some readers. For example, in General Relativity an observer is a future-pointing normalized timelike curve (see for example, Ref. \ref{SW}). For these readers, the connotations of the word ``observation'' would be preferable. (iii) Even though the names ``temporal'' and ``spatial'' for the coordinates are just definitions in P1,  this will suggest  a mathematical formalization of the fact: ``an observer uses four coordinates to describe spacetime; one coordinate for time and the other three for space''. The difference between the temporal and spatial coordinates is stressed in P2 (see also points 2 and 3 in Section 5). (iv) The normalization of the charts (each one centered at $p$) is not restrictive and will be useful. However, one can think that a  physical observer responsible of the coordinate chart will describe the curve $s\mapsto\Phi^{-1}(s,0,0,0)$ (``observer's trajectory''). (v) Throughout this article, differentiability will mean $C^1$, even though there is no problem for accepting $C^2$ or even $C^{\infty}$ (usual discussions on the use of differentiable elements are applicable, see for example, the comments to axiom $PL$ in Ref. \ref{Eh}, p. 27).
(vi) The content of the assertion ``$M$ is a differentiable manifold'' is purely local. However, we will assume, as usual, natural global topological properties, frequently implicit in the definition of manifold: $M$ will be Hausdorff and connected\footnote{Second axiom of numerability can also be assumed. Nevertheless, recall that this axiom can be deduced a posteriori for many interesting cases; for example, when the (connected) manifold admits a non-degenerate metric, Ref. \ref{M} and Ref. \ref{Sp}, p.8-52s.}.

For any observer $O$ around $p$, the tangent vector $\partial_t|_p$ (resp. $\partial_{x^i}|_p$) is the {\em instantaneous temporal unit} or, simply, {\em temporal unit} (resp. {\em i--th spatial unit} $i\in\{1,2,3\}$) of the observer $O$ at $p$. The {\em temporal axis} (resp. {\em i--th spatial axis}) is the subspace of tangent space at $p$, $T_pM$, spanned by the temporal unit (resp. i--th spatial unit). The three spatial axes span the observer's {\em space at rest} $\langle\partial_{x^1}|_p,\partial_{x^2}|_p, \partial_{x^3}|_p\rangle$. The linear form $dt|_p$ will be its {\em clock}.

Given any observer $O$ around $p$,  the basis of $T_pM$, given by its units, $B_p=(\partial_t|_p , \partial_{x^1}|_p, \partial_{x^2}|_p, \partial_{x^3}|_p)$ is the {\em instantaneous observer at $p$} (or simple, {\em observer at $p$})
associated to $O$. Of course, two different observers $O, O'$ around $p$ can yield the same (instantaneous) observer at $p$.

In what follows, only the properties of instantaneous observers will be truly relevant. So, we can use the terms observer ``around $p$'' and ``at $p$'' interchangeably. In fact, from a purely formal viewpoint we could have introduced observers at $p$ directly as an alternative definition in P1, without mentioning observers around $p$ (``...For any $p\in M$ any basis of $T_pM$ will be called {\em (instantaneous) observer at $p$};  the first element of this basis is {\em the temporal unit} of the observer, and the other three are  the {\em spatial units}''). Nevertheless, we think that our choice is clearer from a physical viewpoint. 

\subsection{Second postulate, P2}
As we have already pointed out, the specific method to take coordinates used by ``observers'' is not relevant for P1. But the language above introduced, suggests that each observer may think that uses  ``good instruments'' to measure (at least around the centered point $p$): the coordinate $t$ measures units of time (thus, at least infinitesimally $\partial_t |_p$ is the unit of time), and analogously for the spatial coordinates. Nevertheless, our following postulate P2 will assert that not all the coordinates around $p$ will be equally good. In plane words, a principle of ``restricted democracy'' will be stated:

\begin{itemize}
\item Around each event $p$, there will be  a set  of  observers $S_p$ which are priviledged, at least in an idealized limit  at $p$ (infinitesimally, in the properties relevant to the corresponding instantaneous observers).

\item But these observers are not priviledged among them. That is,  when they compare their measures of time (and, independently, of space) at $p$, no one of the observers can be priviledged on the other. For us, this is the only sensible definition of the physical intuitive idea {\em two observers measure around an event by using the same ``ideal'' or ``fundamental'' units of time and length}.
\end{itemize}

Of course, this principle is satisfied by Newton's theory (standard = inertial, and the ``infinitesimal limit'' is not necessary) and Einstein's General Relativity (infinitesimal standard observer at $p$= orthonormal basis of tangent space at $p$; one can find such observers among ``freely falling observers''). And it seems reasonable also for any other conceivable theory of spacetime, at least in the classical (non-quantum) case.\\

\noindent{\bf P2 (Standard observers)} For each $p\in M$ there exists a (non-empty) set, $S_p$, of distinguished observers around $p$, {\em compatible with a fundamental system of units at $p$} (or simply, {\em compatible}). This means that $\forall O,\tilde O\in S_p$, with $O\equiv(U,\Phi)=(t,x^1,x^2,x^3)$ and $\tilde O\equiv(\tilde U, \tilde\Phi)=(\tilde t,{\tilde x}^1,{\tilde x}^2,{\tilde x}^3)$:
$$
\partial_{\,t}\,{\tilde t}|_p=\partial_{\,{\tilde t}}\,t|_p\,\;\,and\,\;\,
 \partial_{\,x^j}{\tilde x}^i|_p=\partial_{\,{\tilde x}^i}x^j|_p,\,\;
  \forall i,j\in\{1,2,3\}.
$$
Such observers will be called {\em standard observers}.\\

\noindent{\em Note.} The real numbers $ \partial_{\,t}\,{\tilde t}|_p, \partial_{\,{\tilde t}}\,t|_p\partial_{\,x^j}{\tilde x}^i|_p, \partial_{\,{\tilde x}^i}x^j|_p $ can be seen as components of the transition matrix between the instantaneous observers at $p$ associated to $O$ and $\tilde O$ (see Section 3).  Thus, the question whether  $O, \tilde O$ are compatible with a fundamental system of units at $p$, depends exclusively on the instantaneous associated observers. $S_p^i$ will denote the set of (instantaneous) standard observers at $p$ (those instantaneous observers associated to observers in $S_p$).\\

\noindent{\em Discussion {\rm 2}}. According to {\rm P2}, it is possible to choose experimentally a set of observers at each $p$ verifying the previous symmetry conditions between their partial derivatives. However, {\rm P2} does not inform about the specific experimental selecting method. Symmetry conditions at $p$, only mean that:

\begin{enumerate}
\item The ${\tilde O}$ time, measured with the $O$ clock, goes by as the $O$ time, observed by the ${\tilde O}$ clock ($O$ and ${\tilde O}$, measure at $p$ using the same ``fundamental'' unit of time).

\item For all $i,j\in\{1,2,3\}$, the $i$--th spatial unit of ${\tilde O}$, measured with the $j$--th rod of $O$, is identical to the $j$--th spatial unit of $O$, observed with the $i$--th rod of $\tilde O$ ($O$ and ${\tilde O}$, measure at $p$ using the same ``fundamental'' unit of length).
\end{enumerate}
Note that when the spaces at rest of $O$ and ${\tilde O}$ coincide, the last condition is equivalent to something which is very familiar: {\em if one takes the Euclidean metric for which the spatial units of $O$ form an orthonormal basis, then the spatial units of ${\tilde O}$ also form an orthonormal basis}\footnote{In particular, both observers can agree when they measure ``lengths'' at $p$. More precisely: let  $v\in T_pM$ be a tangent vector which belongs to the common rest space of $O$ and $\tilde O$ at $p$. Writing $v= \sum_i a^i \partial_{x^i}\mid _p
= \sum_j \tilde a^j \partial_{\bar x^j}\mid _p$ the common number 
$(\sum_i (a^i)^2)^{1/2}=(\sum_j(\tilde a^j)^2)^{1/2}$ can be called ``length of $v$'' by both observers. (An analogous agreement would be possible if $O, \tilde O$ has equal temporal axis, and $v$ belong to this axis.)}. And when the rest spaces do not coincide, the extrapolation of this interpretation would oblige the substitution of the spatial units of ${\tilde O}$ by its projection onto the space at rest of $O$ (in the direction of the $O$ temporal axis). Recall that we do not even impose on this projection the restriction to yield linearly independent vectors, we only assume that the symmetry holds.

Recall that P2 is valid for both, Einstein and Newton theories. And it would be also true for any other more exotic theory of spacetime, under our two minimal assumptions, i.e.: (i) 1+3 coordinates are needed, and (ii) a final agreement between the ``best observers'' (those which use the best possible ideal instruments for the theory) is possible. Let us  explain this  with an example. 
Let  $O$, $O'$ be two observers of spacetime $M$ (in the sense of P1) around an event $p\in M$. Traditionally, the assertion ``$O$ and $O'$ measure with ideal instruments at $p$'' has different meaning according to the geometrical structure that, finally, one considers on $M$:

\begin{itemize}

\item[(a)] For classical Newtonian theory, this assertion means: (i) both, the clock of $O$ and the clock of $O'$ measure the {\em absolute time}\footnote{And, thus, generate Leibnizian absolute clock 1-form in Section 4; see also Section 5, point~3.}, and (ii) both, the three spatial yardsticks of $O$ and the three spatial yardsticks of $O'$, generate orthonormal bases of  Euclidean space. 
\item[(b)] For Special and General Relativity, the assertion means: both, the clock and the three yardsticks of $O$ and the clock and the three yardsticks of $O'$, 
generate an  orthonormal basis of a Lorentzian metric at $p$.
\end{itemize}
Nevertheless, P2 shows that there exist an infinitesimal symmetry at $p$ which permits $O$ and $O'$ to distinguish if they use ideal instruments, and this symmetry can be defined {\em independently of the geometrical model of spacetime $M$}. In fact, it is previous to any geometrical model, and it represents a foundation for the possible geometrical models.

\subsection{Third postulate, P3}

Standard observers are defined around each individual event $p$, and our last postulate will be just a natural assumption on differentiability when $p$ varies. Nevertheless, this has some technical difficulties. On one hand, our language ``standard observer'' (``around'' or ``at'' $p$) does not suggest directly a concept of differentiability. On the other hand, even though we need a concept of differentiability through $p$, we do not want to be unnecesarily restrictive. Especially, we do not want to impose at the same time an assumption on differentiability at each $p$, for each single $S_p$. Recall that $S_p$ (as well as the corresponding set of instantaneous observers $S_p^i$) will be determined at each event  by an experimental or theoretical method, and there is no reason for $S_p, S_p^i$ to have any structure ``a priori''. For example, the existence of a priviledged reference system at an event $p$ or, say, the finiteness of $S_p$ (and, thus, $S_p^i$), are admissible. But perhaps at a different event $q$ there is a sort of symmetry which makes reasonable to assume that $S_q$ and  $S_q^i$ are infinite. Cases like this must not be excluded by our assumption of differentiability. In particular, not too standard proposals, as the one in Ref. \ref{EM} and Ref. \ref{ZTE}, will fit under our approach. However, in order to solve these technical difficulties, we will go further in the question what a ``fundamental system of units'' of length and time is.

Let $L_pM$ be the collection of all the (ordered) bases (or linear frames) on $T_pM$, and let $LM= \cup_{p\in M} L_pM$ be the fiber bundle of the bases on $M$. It is well-known that $L_pM$ and $LM$ admit canonical structures of differentiable manifolds. Clearly, the collection of instantaneous standard observers $S^i_p$ at $p$ is a subset of  the manifold $L_pM$. We can wonder if this subset $S_p^i$ can be seen as a (embedded) submanifold\footnote{Recall that when a subset of a manifold is a embedded submanifold then it has the induced topology and its differentiable structure is unique (see for example, Ref. \ref{Wa}, p. 27).}. But, as commented above,  the answer to this question is negative, in principle. Nevertheless, $S_p^i$ (or, equivalently, $S_p$) yields naturally a bigger set of basis, {\em the fundamental system of units at $p$}, $S_p^{*i}$, with a deeper physical meaning. And we will prove that $S_p^{*i}$ is a submanifold of $L_pM$.

Consider sets  $S_p^{\alpha}$ of observers around $p$ which verify: (i) $S_p \subseteq S_p^{\alpha}$, (ii) all the observers in $S_p^{\alpha}$ are compatible between them (not only with the observers in $S_p$) and (iii) each $S_p^{\alpha}$ is {\em maximal}, that is, $S_p^{\alpha}$ is not strictly included in another set of observers satisfying (i) and (ii)\footnote{Notice that, given to such sets $S_p^{\alpha}, S_p^{\beta}$, an observer in $S_p^{\alpha}$ may be non-compatible with an observer in $ S_p^{\beta}$. That is, the set $S_p^{com}$ of all the observers around $p$ which are compatible with all the observers in $S_p$, maybe a non-compatible set: two observers $O, O' \in S^{com}$ may be non-compatible between them (necessarily, then $O, O' \not\in S_p)$ .}.

Each set of observers $S_p^{\alpha}$ is not unequivocallly associated to $S_p$; perhaps there are more than one set satisfying (i), (ii), and (iii). But the intersection of all these sets $S_p^{\alpha}$ {\em can be canonically associated to } $S_p$. This intersection,  ${S}^*_p = \cap_{\alpha} {S}_p^{\alpha}$ will be called
 the set of observers around $p$ {\em physically equivalent} to observers $S_p$, and:\\

\noindent {\bf Definition 2.1.} Let $S_p$ be a set of  observers around $p$ compatible with a system of fundamental units in the sense defined in P2, and $S^*_p$ the corresponding set of equivalent observers. The set ${S}^{*i}_p$ of infinitesimal observers at $p$ associated to $S^*_p$ is the  {\em fundamental system of units (of time and length) at $p$} associated to~$S_p$.\\

\noindent {\em Discussion} 3A. As commented above, the transition from ${S}_p$ to ${S}^*_p$ and ${S}^{*i}_p$ is necessary from a mathematical viewpoint, but only because of a technical question of differentiability. Thus, Definition 2.1 as well as previous discussion can be regarded  as a convenient mathematical construction. Nevertheless, all this is also {\em natural from a physical viewpoint}. Recall that, from an experimental physical viewpoint, P2 does not mean that one has a real physical standard observer  around each event of spacetime (as well as the axioms in EPS does not mean that any event is truly crossed by infinitely many real physical freely falling particles or light rays). P2 just expresses the possibility (experimental or theoretical) of finding such observers.

But, once some standard observers around $p$ are identified, and the set  $S_p$  is to be constructed, one must admit, at least from a theoretical viewpoint that,  given $S_p$, {\em all the observers in $S_p^*$  are as good as the observers in $S_p$} (they satisfy the same symmetries). As $S_p^*$ cannot be unequivocally extended to a bigger set of observers, the observers in $S_p^*$ (and only them) can be regarded as physically equivalent to those in $S_p$. The final step, that is, to define $S_p^{*i}$ as the fundamental system of units just expresses: (a) for P2 only the properties of the instantaneous observers are relevant and (b) as  the operational use of the same system of units by two observers was defined in P2, the system of units itself can be defined by means of the observers who use it.

Now, we can come back to the problem of differentiability. Recall that the fundamental system of units ${S}^{*i}_p$ is a subset of $L_pM$, and let $S^{*i}= \cup_{p\in M}S_p^{*i}$ be the collection of all the fundamental units  at every point. As explained above, we will prove (except at most in some residual cases of scarce physical interest) that $S_p^{*i}$ has a natural differentiable structure, that is, $S_p^{*i}$ is a embedded submanifold of $L_pM$. But in this case, it is natural to assume that the  differentiable structures at different points unite in a whole differentiable structure $S^{*i}$; this assumption will be the meaning of our third postulate. In order to cover the residual cases too, the assumption will be imposed only on a suitable open subset $U$.\\

\noindent{\bf P3 (Differentiability)}. The transition between the collections of  
fundamental systems of units at different points $p\in M$ is smooth in a natural sense, that is: if (one shows that, necessarily) $S^{*i}_p$ has a natural differentiable structure (as an embedded submanifold of the space of linear frames $L_pM$), for all p in an open subset $U$ of M, then the collection of all the fundamental units  at every point of $U$, $S_U^{*i}= \cup_{p\in U} S^{*i}_p$ has a natural differentiable structure (as a embedded submanifold of the whole bundle of linear frames $LM$).\\

\noindent{\em Discussion {\rm 3B}}. (i) In fact,  we will see in the Section 3 that (except in the ``residual'' cases) $S_p^{*i}$ not only is a submanifold in $L_pM$ but also satisfies a stronger property: a closed group of matrices acts transitively on $S_p^{*i}$, yielding a natural mathematical structure (Euclidean, Lorentzian, ``Leibnizian'' or ``anti-Leibnizian'') on $T_pM$. Of course, it is completely natural to assume  that this structure varies smoothly from one point to another, and this is equivalent to P3.\footnote{Because of the gap between macroscopic and microscopic physics, smooth assumptions are known to be finally controversial 
(Discussion 1(v) of P1). Summing up,  our way of measuring (or of handling our measures) makes differentiability  mathematically convenient (even unavoidable); nevertheless, it is really physically meaningless. Recall that, in principle, we need just differentiability $C^1$.} (ii) Recall that ``differentiable'', in 
the mathematical formalization of P3, means strictly just the usual concept of $C^1$ differentiability. This sense of differentiability excludes, intuitively, singularities as peaks or the splitting of a curve in two (the curve would not be differentiable in the splitting point).  Nevertheless, in the ``residual cases'', the set $S^{*i}_p$ might have a not too good mathematical structure to apply P3. As we will see, this residual cases must be regarded as a mathematical curiosity, more than as a true physical possibility (for example, they cannot appear if any of the ``optional postulates''  in Section 5 hold). Thus, the ``natural sense'' for differentiability stated in P3 will be enough for our purposes, and one can assume 
$U=M$ in P3 . At any case, we will discuss widely the residual cases and show that they can be controlled completely (including discussions on differentiability, under  more general assumptions than in P3, see Section 4).\\

\noindent {\em Note.} From a practical viewpoint, $S_p$ yields directly the other subsets $S_p^*, S_p^i$ and  $S_p^{*i}$. Moreover, even though the distinction {\em a priori} of $S_p$ and $S_p^*$ is important,  we can see now that there is no problem if one assumes finally $S_p=S_p^*$ (resp. $S_p^i =S_p^{*i}$). In spite of this, the superscript, $^{(*)}$, will be maintained because of the conceptual distinction between $S_p$ and $S_p^*$. But, in what follows, our notation will be simplified suppressing the superscript, $^{(i)}$, and identifying observers at and around $p$, if there is no possibility of confusion (all the observers around $p$ which yields the same instantaneous observer at $p$ can be identified with this instantaneous observer; each observer around $p$ will also be regarded as a basis of $T_pM$; only the properties of these basis will be relevant).

\section{Structures induced in the tangent space by standard observers}

The purpose of this section is to obtain Theorem 3.1, which is a purely mathematical result on the structure of each $S_p (\equiv S_p^i)$, deduced from P2. Even though we use rather elementary mathematical tools, the complete process may be somewhat misleading. To avoid to get lost, we give a brief summary at the end (complemented with Remark 3.1 and Definition 3.1).

Let us introduce the following notation. For any two observers at $p\in M$, $O,{\tilde O}$, let $M(Id,O\leftarrow {\tilde O})$ be the transition matrix from ${\tilde O}$ to $O$; that is, if $M(Id,O\leftarrow {\tilde O})$ is multiplied to the right by the column-coordinates of a vector $v\in T_pM$ in the basis ${\tilde O}$, we obtain the column-coordinates of the same vector $v$ in $O$. Each regular (real) matrix $A=(a_{\mu\nu})\in Gl(4,\mathbf{R})$ will be divided into boxes as follows:

\[\left(
\begin{array}{c|c}
 a_{00} & {\mathbf a_h} \\ \hline
^t{\mathbf a_v}   & {\mathbf{\hat A}}
\end{array}
\right),
\]
where we define, ${\mathbf a_h}=\pmatrix{a_{01} & a_{02} & a_{03}}$, ${\mathbf a_v}=\pmatrix{
a_{10} & a_{20} & a_{30}}$, the superscript $^{(t)}$ denotes transpose, and:

\[
{\mathbf{\mathbf{\hat A}}}=
\pmatrix{
 a_{11} & a_{12} & a_{13} \cr
 a_{21} & a_{22} & a_{23} \cr
 a_{31} & a_{32} & a_{33} \cr
        }.
\]
Moreover, $\|\cdot\|$ will denote the canonical Euclidean norm of $\mathbf{R}^3$, $O(n,\mathbf{R})$ the orthogonal group of order $n$, and $\{\pm1\}\times O(3,\mathbf{R})$, the subgroup of $O(4,\mathbf{R})$ which is equal to the product of $\{-1,1\}$ and $O(3,\mathbf{R})$ ($A\in\{\pm1\}\times O(3,\mathbf{R})$ if and only if $a_{00}^2=1$, ${\mathbf a_h}=\mathbf 0={\mathbf a_v}$ and ${\mathbf{\hat A}}\in O(3,\mathbf{R})$). The following two lemmas collect the  algebraic properties of the matrices which can be deduced directly from {\rm P2}.\\

\noindent {\bf Lemma 3.1.} Let $S_p$ be a compatible set at a point $p\in M$ and let $O,{\tilde O}\in S_p$, be with $A=M(Id,O\leftarrow{\tilde O})$. Then:

\begin{enumerate}
\item For ${\mathbf{\tilde a}_h}=\pmatrix{{\tilde a}_{01} & {\tilde a}_{02} & {\tilde a}_{03}},{\mathbf{\tilde a}_v}=\pmatrix{{\tilde a}_{10} & {\tilde a}_{20} & {\tilde a}_{30}}\in\mathbf{R}^3$:
\begin{equation}
A^{-1} = M(Id,{\tilde O}\leftarrow O)=\left(
\begin{array}{c|c}
 a_{00} & {\mathbf{\tilde a}_h} \\ \hline
^t{\mathbf{\tilde a}_v}  & ^t{\mathbf{\hat A}}
\end{array}
\right)
\end{equation}
\noindent Moreover:
\begin{enumerate}
\item[(i)]  
${\mathbf a_h}\not=\mathbf 0$ (resp. ${\mathbf a_v}\not=\mathbf 0$) if and only if ${\mathbf{\tilde a}_h}\not=\mathbf 0$ (resp. ${\mathbf{\tilde a}_v}\not=\mathbf 0$).
\item[(ii)]
if $a_{00}\not=0$, then $det^2A=1$; if $a_{00}=0$ then ${\mathbf a_h}\not=
 \mathbf 0\not={\mathbf a_v}$ and $rank\,{\mathbf{\hat A}}=2$.
\end{enumerate}
\item Assume that ${\mathbf a_h}\not=\mathbf 0$, then there exist ${\tilde k},k\in\mathbf{R}$ such that:  
\begin{enumerate}
\item[(i)] ${\mathbf a_v}=k\,{\mathbf{\tilde a}_h}$ and ${\mathbf{\tilde a}_v}={\tilde k}\,{\mathbf a_h}$, with: 
\begin{equation}
\label{k}
k\,\|{\mathbf{\tilde a}_h}\|^2=1-a_{00}^2={\tilde k}\,\|{\mathbf a_h}\|^2.
\end{equation}
\item[(ii)] $k={\tilde k}\,det^2A$. Moreover, ${\tilde k}\not=k$ if and only if $det^2A\not=1$; in this case, ${\tilde k}$ and $k$ are positive.
\end{enumerate}
\end{enumerate}
{\rm Proof}: First, (\rm 1) is just to apply the definition of compatibility for $O,{\tilde O}$. On the other hand, the product of matrices by boxes $A\,A^{-1}=I_4=A^{-1}\,A$ implies the relations:
\begin{equation}
{\mathbf a_h}\,^t{\mathbf{\tilde a}_v}=1-a_{00}^2={\mathbf{\tilde a}_h}\,^t{\mathbf a_v} 
\end{equation}
\begin{equation}
a_{00}\,{\mathbf{\tilde a}_h}+{\mathbf a_h}\,^t{\mathbf{\hat A}}=\mathbf 0=a_{00}\,{\mathbf a_h}+{\mathbf{\tilde a}_h}\,{\mathbf{\hat A}} 
\end{equation}
\begin{equation} 
a_{00}\,{\mathbf a_v}+{\mathbf{\tilde a}_v}\,^t{\mathbf{\hat A}}=\mathbf 0=a_{00}\,{\mathbf{\tilde a}_v}+{\mathbf a_v}\,{\mathbf{\hat A}}
\end{equation}
\begin{equation}
^t{\mathbf a_v}\,{\mathbf{\tilde a}_h}+{\mathbf{\hat A}}\,^t{\mathbf{\hat A}}=I_3=\,^t{\mathbf{\tilde a}_v}\,{\mathbf a_h}+\,^t{\mathbf{\hat A}}\,{\mathbf{\hat A}}.
\end{equation}
Clearly, {\rm 1(i)} is immediate from (\rm 3),(\rm 4) and (\rm 5). For the first implication {\rm 1(ii)}, recall that, from the elemental algorithm to calculate the inverse matrix:
\begin{equation}
\frac{det\,{\mathbf{\hat A}}}{detA}=a_{00}=\frac{det\,^t{\mathbf{\hat A}}}{det A^{-1}}\;,
\end{equation}
therefore, $a_{00}\not=0\Rightarrow det\,{\mathbf{\hat A}}\,(= det\,^t{\mathbf{\hat A}})\not = 0$, and the result is straightforward applying ({\rm 7}) again. The second is obvious from ({\rm 3}), ({\rm 7}) and the regularity of $A$. For {\rm 2(i)}, recall first that equation (\rm 2) is a consequence of the two first equalities in {\rm 2(i)} and of (\rm 3); so, we just have to prove these equalities {\rm 2(i)}. From (\rm 6):
\[
^t{\mathbf a_v}\,{\mathbf{\tilde a}_h}=I_3-{\mathbf{\hat A}}\,^t{\mathbf{\hat A}}=\,^t(I_3-{\mathbf{\hat A}}\,^t{\mathbf{\hat A}})=\,^t{\mathbf{\tilde a}_h}\,{\mathbf a_v},
\]
that is, the decomposable matrix $^t{\mathbf a_v}\,{\mathbf{\tilde a}_h}$ is symmetric. Thus, the vectors $\{{\mathbf a_v},{\mathbf{\tilde a}_h}\}$ are linearly dependent, and the first equality in {\rm 2(i)} is obtained (the second one is deduced analogously).

For the first assertion in the case {\rm 2(ii)}, using again the algorithm to calculate the inverse matrix:
\[{\tilde a}_{10}=-\frac{det\pmatrix{
 k\,{\tilde a}_{01} & a_{12} & a_{13}\cr
 k\,{\tilde a}_{02} & a_{22} & a_{23}      \cr
 k\,{\tilde a}_{03} & a_{32} & a_{33}}
                 }{det A}=-\frac{k}{det^2A}\frac{
                  det\pmatrix{
 {\tilde a}_{01} & {\tilde a}_{02} & {\tilde a}_{03} \cr
 a_{12}   & a_{22}   & a_{32}   \cr
 a_{13}   & a_{23}   & a_{33}}
                                                }{det A^{-1}}
\]
\[=\frac{k}{det^2 A}\,a_{01},
\]
and analogously for ${\tilde a}_{20},{\tilde a}_{30}$, so:
\[
\left(
 {\tilde k}-\frac{k}{det^2A}
\right){\mathbf a_h}=\mathbf 0,\;\;\;\;\hbox{thus}\;\;\;\;{\tilde k}=\frac{k}{det^2A}.
\]
This last equality (with (\rm 2) and {\rm 1(ii)} for the case $k=0={\tilde k}$) ends the proof.

\begin{flushright}
$\triangleleft$
\end{flushright}

The relation between the timelike units of $O$ and ${\tilde O}$ is given by:
\[
\partial_{t}|_p-a_{00}\,\partial_{{\tilde t}}|_p=\sum_{i=1}^3{\tilde a}_{i0}\,\partial_{{\tilde x}^i}|_p
 \;\;\;\;\;and\;\;\;\;\;
  \partial_{{\tilde t}}|_p-a_{00}\,\partial_{t}|_p=\sum_{j=1}^3a_{j0}\,       \partial_{x^j}|_p\,,
\]
so, the timelike unit of $O$ lies in the local rest space of ${\tilde O}$ (with coordinates ${\mathbf{\tilde a}_v}$), if and only if $a_{00}=0$; in this case, the timelike unit of ${\tilde O}$ also lies in the local rest space of $O$ (with coordinates ${\mathbf a_v}$).

On the other hand, when a third compatible observer is taken into account, the corresponding $k,{\tilde k}$ must be related.\\

\noindent {\bf Lemma 3.2.} Fixed $p\in M$, let $O^{(\alpha)}$, $O^{(\beta)}$ and ${\tilde O}$ be three observers in a compatible set $S_p$, and put $A=M(Id,O^{(\alpha)}\leftarrow{\tilde O})$, $B=M(Id,O^{(\beta)}\leftarrow{\tilde O})$. Then:
\begin{equation}
{\mathbf a_h}\,^t{\mathbf{\tilde b}_v}={\mathbf b_h}\,^t{\mathbf{\tilde a}_v}\,\;\;\,and\,\;\;\,^t{\mathbf{\tilde a}_h}\,{\mathbf b_v}=\,^t{\mathbf a_v}\,{\mathbf{\tilde b}_h}.
\end{equation}
Thus, if there are  $k_\alpha,{\tilde k}_\alpha,k_\beta,{\tilde k}_\beta\in\mathbf{R}$, such that:
\[
{\mathbf a_v}=k_\alpha\,{\mathbf{\tilde a}_h},\,{\mathbf{\tilde a}_v}={\tilde k}_\alpha\,{\mathbf a_h},{\mathbf b_v}=k_\beta\,{\mathbf{\tilde b}_h}
 \,\;\,and\,\;\,{\mathbf{\tilde b}_v}={\tilde k}_\beta\,{\mathbf b_h},
\]
necessarily:
\[
{\tilde k}_\beta\,{\mathbf a_h}\,^t{\mathbf b_h}={\tilde k}_\alpha\,{\mathbf b_h}\,^t{\mathbf a_h}
 \,\;\;\,and\,\;\;\,k_\beta\,^t{\mathbf{\tilde a}_h}\,{\mathbf{\tilde b}_h}=k_\alpha\,^t{\mathbf{\tilde a}_h}\,{\mathbf{\tilde b}_h}.
\]
Therefore:
\begin{equation}
{\mathbf a_h}\,^t{\mathbf b_h}(={\mathbf b_h}\,^t{\mathbf a_h})\not=0\Rightarrow {\tilde k}_\beta={\tilde k}_\alpha
 \,\;\;\,and\,\;\;\,{\mathbf a_h}\not=\mathbf 0\not={\mathbf b_h}\Rightarrow k_\beta=k_\alpha.
\end{equation}
\noindent {\rm Proof}: Putting $C=M(Id,O^{(\beta)}\leftarrow O^{(\alpha)})$, all the assertions are proven just taking into account the product of matrices by boxes $B\,A^{-1}=C$ and $C^{-1}=A\,B^{-1}$. In fact, on one hand:
\[
{\mathbf b_h}\,^t{\mathbf{\tilde a}_v}=c_{00}-a_{00}\,b_{00}={\mathbf a_h}\,^t{\mathbf{\tilde b}_v},
\]
and on the other:
\[
^t{\mathbf a_v}\,{\mathbf{\tilde b}_h}=\,^t{\mathbf{\hat C}}-\, {\mathbf{\hat A}}\,^t{\mathbf{\hat B}}=\,^t({\mathbf{\hat C}}-{\mathbf{\hat B}}\,^t{\mathbf{\hat A}})=\,^t(^t{\mathbf b_v}{\mathbf{\tilde a}_h})= \,^t{\mathbf{\tilde a}_h}\,{\mathbf b_v},
\]
as required.

\begin{flushright}
$\triangleleft$
\end{flushright}

Next, we will consider the relation of equivalence $\sim$ in any compatible set $S_p$ given by:
\[
O\sim{\tilde O}\,\;\;\,\Leftrightarrow \,\;\;\,det^2M(Id,O\leftarrow{\tilde O})=1.
\]
A class of equivalence ${\cal C}_p$ of the quotient set $S_p/\sim$ is called {\rm proper}, if:
\[
\exists O',{\tilde O}'\in {\cal C}_p:\,M(Id,O'\leftarrow{\tilde O}')\not\in \{\pm1\}\times O(3,\mathbf{R})
\]
(otherwise, the class is called {\rm improper}). Let the square matrix
\[
I_3^{(k)}=
\left(
 \begin{array}{c|c}
 k             & \mathbf 0 \\ \hline
 ^t\mathbf 0   & I_3
\end{array}
\right),\,\;\,\forall k\in\mathbf{R}, 
\]
be, and let $S^1$ be the circle obtained by identifying the two extremes $\pm \infty$ in the set of the extended real numbers $\mathbf{R}^* =[-\infty,+\infty]$  to an only point $\omega$. A regular matrix $A\in Gl(4,\mathbf{R})$ is called $k$-{\rm congruent} for $k\in S^1\backslash\{0, \omega\}$, if and only if:
\begin{equation}
^tA\,I_3^{(k)}\,A = I_3^{(k)} 
\end{equation}
or, equivalently, if $A\,I_3^{(1/k)}\,^tA = I_3^{(1/k)}$. Taking into account that in this case $det^2A=1$, this definition can be extended in a natural way to the cases $k=0,\,\omega$. That is, given $A\in Gl(4,\mathbf{R})$, $A$ is $0$-{\rm congruent}, if:
\begin{equation}
 ^tA\,I_3^{(0)}\,A = I_3^{(0)} \;\;\;\;\;and\;\;\;\;\;det^2A =1,
\end{equation}
and $A$ is $\omega$-{\rm congruent}, if:
\[
 A\,I_3^{(0)}\,^tA = I_3^{(0)}
  \;\;\;\;\;and\;\;\;\;\;det^2A =1. 
\]
At last, we define for each $k\in S^1$:
\[
O^{(k)}(4,\mathbf{R})=\{A\in Gl(4,\mathbf{R})\,/\,\mbox{$A$ is $k$-congruent}\},
\]
which is a subgroup of $Gl(4,\mathbf{R})$.

From these definitions, Lemma {\rm 3.1} and (\rm 3), it is straightforward to check that $A= M(Id,O\leftarrow{\tilde O})$ is $k$-congruent, $k\in\mathbf{R}$, if and only if, ${\mathbf a_v}=k\,{\mathbf{\tilde a}_h}$ and ${\mathbf{\tilde a}_v}=k\,{\mathbf a_h}$; and $A$ is $\omega$-congruent if and only if  ${\mathbf a_h} = 0$. Two  observers $O,{\tilde O}\in S_p$ are said \emph{congruent}, if $M(Id,O\leftarrow{\tilde O})$ is $k-$congruent for some $k\in S^1$, that is, if $det^2M(Id,O\leftarrow{\tilde O})=1$; otherwise, they are \emph{incongruent}. Obviously, if $A\in Gl(4,\mathbf{R})$ is the  transition matrix between two incongruent observers (${\mathbf a_h}\not=\mathbf 0$, ${\mathbf a_v}=k\, {\mathbf{\tilde a}_h}$, ${\mathbf{\tilde a}_v}={\tilde k}\,{\mathbf a_h}$ and  $0<{\tilde k}\not=k>0$) then $A^2$ is not the transition matrix between any pair of standard observers (otherwise, from ${\tilde k}\,^t{\mathbf a_h}\,{\mathbf{\tilde a}_h}+\,^t{\mathbf{\hat A}}^2=\,^t(k\,^t{\mathbf{\tilde a}_h}\,{\mathbf a_h}+{\mathbf{\hat A}}^2)$ we would obtain ${\tilde k}=k$).\\   

\noindent {\bf Lemma 3.3.} For any compatible set $S_p$:

\begin{enumerate}
\item 
If all the elements of $S_p$ are congruent then either $M(Id,O\leftarrow {\tilde O})\in\{\pm1\}\times O(3,\mathbf{R})$, $\forall O,{\tilde O}\in S_p$ or there exists a unique $k\in S^1$ such that $A=M(Id,O\leftarrow{\tilde O})\in O^{(k)}(4,\mathbf{R})$, $\forall O,{\tilde O}\in S_p$.
\item 
If $S_p$ contains incongruent observers then each ${\tilde O}\in S_p$ fixes a unique  $k\in\mathbf{R}$, which must be positive, such that $\forall O\in S_p$, if $A=M(Id,O\leftarrow{\tilde O})$ then ${\mathbf a_v}=k\,{\mathbf{\tilde a}_h}$; in particular, ${\mathbf a_v}=0\Leftrightarrow{\mathbf a_h}=\mathbf 0\Leftrightarrow A\in\{\pm1\}\times O(3,\mathbf{R})$. As a consequence: if ${\tilde O}_1,{\tilde O}_2$ determine $k_1,k_2>0$ then ${\tilde O}_2\sim{\tilde O}_1\Leftrightarrow k_2=k_1\Leftrightarrow M(Id,{\tilde O}_1\leftarrow{\tilde O}_2)\in O^{(k)}(4,\mathbf{R})$, for $k=k_2=k_1$.
\end{enumerate}

\noindent {\rm Proof}: Assume first that there are no incongruent observers. If, in addition, $M(Id,O\leftarrow{\tilde O})\in O^{(\omega)}(4,\mathbf{R})$, $\forall O,{\tilde O}\in S_p$ and we assume that one of these matrices $C$ is congruent for some other $k\in\mathbf{R}$, we deduce from formulas (\rm 10) and (\rm 11):
\[\left(
\begin{array}{c|c}
 \pm k & {\mathbf c_v} \\ \hline
 \mathbf 0    & ^t{\mathbf{\hat C}}
\end{array}
\right)=\left(
\begin{array}{c|c}
 \pm 1 & {\mathbf c_v} \\ \hline
 \mathbf 0    & ^t{\mathbf{\hat C}}
\end{array}
\right)\left(
\begin{array}{c|c}
 k  & \mathbf 0 \\ \hline
 \mathbf 0 & I_3
\end{array}
\right)
\]
\[
=\left(
\begin{array}{c|c}
 k  & \mathbf 0 \\ \hline
 \mathbf 0 & I_3
\end{array}
\right)\left(
\begin{array}{c|c}
 \pm 1  & \mathbf 0 \\ \hline
 ^t\mathbf{\tilde c}_v & ^t{\mathbf{\hat C}}
\end{array}
\right)=\left(
\begin{array}{c|c}
 \pm k  & \mathbf 0 \\ \hline
 ^t\mathbf{\tilde c}_v & ^t{\mathbf{\hat C}}
\end{array}
\right).
\]
Thus ${\mathbf c_v}=\mathbf 0=\mathbf{\tilde c}_v$, that is, all these matrices are elements of  $\{\pm1\}\times O(3,\mathbf{R})$. Otherwise, $k=\omega$ is unequivocally determined.

On the other hand, if at least two standard observers $O^{(0)},{\tilde O}^{(0)}\in S_p$ have ${\mathbf a_h}\not=\mathbf 0$ for $A=M(Id,O^{(0)}\leftarrow {\tilde O}^{(0)})$, then there exist a unique $k\in\mathbf{R}$ such that ${\mathbf a_v}=k\,{\mathbf{\tilde a}_h}$ and ${\mathbf{\tilde a}_v}=k\,{\mathbf a_h}$. Given a new observer $O\in S_p$ and defined $B=M(Id,O\leftarrow{\tilde O}^{(0)})$, then necessarily  ${\mathbf b_v}=k\,{\mathbf{\tilde b}_h}$ (this happens both, when  ${\mathbf b_h}=\mathbf 0$, by applying (\rm 8) to $O^{(0)}$, $O$, ${\tilde O}^{(0)}$, in order to obtain that also ${\mathbf b_v}=\mathbf 0$; and when ${\mathbf b_h}\not=\mathbf 0$ by applying (\rm 9)). That is, $B$ is congruent for the same  $k$ above. As this is also true for any other ${\tilde O}\in S_p$, then $M(Id,O\leftarrow {\tilde O})\in O^{(k)}(4,\mathbf{R})$, $\forall O,{\tilde O}\in S_p$.

Assume now that $S_p$ contains incongruent observers. If ${\tilde O}\in S_p$ then there exists ${\hat O}\in S_p$, with $B=M(Id,{\hat O}\leftarrow{\tilde O})$ and $det^2B\not=1$. As ${\hat O}$ and ${\tilde O}$ are incongruent, ${\mathbf b_h}\not=\mathbf 0$ and there exists a unique real number $k>0$ satisfying ${\mathbf b_v}=k\,{\mathbf{\tilde b}_h}$ ({\rm 1(ii)} and {\rm 2} of Lemma \rm{3.1}). Given any other $O\in S_p$, the matrix $A=M(Id, O\leftarrow{\tilde O})$ also satisfies  ${\mathbf a_v}=k\,{\mathbf{\tilde a}_h}$ (this happens both, when ${\mathbf a_h}=\mathbf 0$, by applying (\rm 8) to $O$, ${\hat O}$, ${\tilde O}$, in order to deduce that also ${\mathbf a_v}=\mathbf 0$ ---which happens if and only if $A\in\{\pm1\}\times O(3,\mathbf{R})$--- and when ${\mathbf a_h}\not=\mathbf 0$, by applying (\rm 9)). Finally, from {\rm 2(ii)} of Lemma {\rm 3.1}, we also have $O\in [{\tilde O}]\Leftrightarrow det^2A=1\Leftrightarrow{\mathbf{\tilde a}_v}=k\,{\mathbf a_h}\Leftrightarrow A\in O^{(k)}(4,\mathbf{R})$.

\begin{flushright}
$\triangleleft$
\end{flushright}

Obviously, if a compatible set contains incongruent observers then it has two or more classes of equivalence; in this case, we have the following.\\

\noindent {\bf Lemma 3.4.} Let $[O^{(i)}],[O^{(j)}]$ be two distinct classes of equivalence in a compatible set $S_p$. Then, the corresponding $k_i, k_j>0$ are different. As a consequence,  fixed ${\tilde O}\in S_p$ and defined $A^{(i)}=M(Id,O^{(i)}\leftarrow {\tilde O})$, $A^{(j)}=M(Id,O^{(j)}\leftarrow{\tilde O})$, we have that ${\mathbf a_h}^{(i)}\,^t{\mathbf a_h}^{(j)}=0$.\\

\noindent {\rm Proof}: The first assertion is a straightforward consequence of {\rm 2(ii)} in Lemma {\rm 3.1}; in fact, recall that if $A=M(Id,O^{(j)} \leftarrow O^{(i)})$ then ${\mathbf a_v}=k_i\,{\mathbf{\tilde a}_h}$, ${\mathbf{\tilde a}_v}=k_j\,{\mathbf a_h}$,  and $O^{(i)}$, $O^{(j)}$ are incongruent ($0<k_j\not=k_i>0$). On the other hand, ${\mathbf{\tilde a}_v}^{(i)}=k_i\,{\mathbf a_h}^{(i)}$ and ${\mathbf{\tilde a}_v}^{(j)}=k_j\,{\mathbf a_h}^{(j)}$; so, (\rm 9) implies the last assertion (otherwise, if ${\mathbf a_h}^{(i)}\,^t{\mathbf a_h}^{(j)} \not=0$, then $k_j=k_i$).

\begin{flushright}
$\triangleleft$
\end{flushright}

Clearly, if ${\cal C}_p$ is a class of equivalence of a compatible set $S_p$, $O,O'\in {\cal C}_p$ and ${\tilde O}\in S_p\backslash{\cal C}_p$, with $A=M(Id,O \leftarrow{\tilde O})$, $B=M(Id,O'\leftarrow{\tilde O})$ and $C= M(Id,O\leftarrow O')$, then:
\begin{equation}
\{{\mathbf a_h},{\mathbf b_h}\}\mbox{ is linearly independent}\,\;\,
 \Leftrightarrow\,\;\,{\mathbf c_h}\not=\mathbf 0.
\end{equation}
In fact: $b_{00}=0=a_{00}$, ${\mathbf b_h}\not=\mathbf 0\not={\mathbf a_h}$, $\mathbf 0={\mathbf b_h}\,^t{\mathbf{\hat B}}$ and ${\mathbf c_h}={\mathbf a_h}\,^t{\mathbf{\hat B}}$ (recall that from Lemma {\rm 3.1}, $b_{00}=0\Rightarrow rank\,{\mathbf{\hat B}}=2$). So, we can determine exactly what happens when there are incongruent observers.\\

\noindent {\bf Proposition 3.1.} If $S_p$ contains  incongruent observers then ${S_p/\sim}$ has two, three or four  classes ${\cal C}_p^{(1)},{\cal C}_p^{(2)},\dots,{\cal C}_p^{(N)}$, and there are no more than $4-N$ proper classes. Moreover, each one of these $N$ classes determines a positive real number $k_1,k_2,\dots,k_N$ in such a way that if $O\in{\cal C}_p^{(i)}$ and ${\tilde O}\in{\cal C}_p^{(j)}$:
\[
^tM(Id,O\leftarrow{\tilde O})\,I_3^{(k_i)}\,M(Id,O\leftarrow{\tilde O})=I_3^{(k_j)},
\]
where $k_j=k_i$ if and only if ${\cal C}_p^{(j)}={\cal C}_p^{(i)}$.\\

\noindent {\rm Proof}: Fixed a class $[{\tilde O}]$, the relations of orthogonality ${\mathbf a_h}^{(i)}\,^t{\mathbf a_h}^{(j)}=0$ in Lemma {\rm 3.4} imply that there are at most other three  distinct classes $[O^{(1)}]$, $[O^{(2)}]$ and $[O^{(3)}]$. Moreover, if  ${S_p/\sim}$ contains  effectively four classes of equivalence then all of them are improper. In fact: if $O\in [{\tilde O}]$ with $A=M(Id,O\leftarrow{\tilde O})$, then from the mentioned relations ${\mathbf a_h}\, ^t{\mathbf a_h}^{(j)}=0$, $\forall j\in\{1,2,3\}$; thus ${\mathbf a_h}=\mathbf 0$ and $A\in\{\pm1\}\times O(3,\mathbf{R})$ (\rm 2 in Lemma {\rm 3.3}).

Now, assume that $S_p\backslash{\cal C}_p$ contains three elements. Then $S_p\backslash{\cal C}_p$ may have either three improper classes or  at least one proper class. In this last case, let $O^{(1)}$, ${\tilde O}^{(1)}$ be two related standard observers, with $M(Id,O^{(1)}\leftarrow {\tilde O}^{(1)})=C$ and ${\mathbf c_h}\not=\mathbf 0$. It is clear that none of the remaining classes $[O^{(2)}]$ and $[O^{(3)}]$, can be proper (by a reasoning as in the previous case and (\rm 12)).

Finally, if  $S_p/\sim$ has just two elements then the relations of orthogonality do not forbid the existence of two proper classes. The remainder is straightforward.

\begin{flushright}
$\triangleleft$
\end{flushright}

As a consequence of this last proposition and {\rm 1} in Lemma {\rm 3.3}, the following complete classification of the sets of compatible observers can be given.\\

\noindent {\bf Theorem 3.1.} Any compatible set $S_p$, satisfies one and only one of the following assertions:
\begin{enumerate}
\item 
{\rm (Regular case)} There exists a unique
 $k\in S^1$ such that: $M(Id,{\tilde O} \leftarrow O)\in O^{(k)}(4,\mathbf{R})$, $\forall O,{\tilde O}\in S_p$.
\item 
{\rm (Degenerate case)} $M(Id,O\leftarrow{\tilde O})\in\{\pm1\}\times O(3,\mathbf{R})$ (or equally, $M(Id,$ $O \leftarrow \tilde O)$ is $k$-congruent, $\forall k\in S^1$), $\forall O,{\tilde O}\in S_p$.
\item 
{\rm (Residual case)} $S_p$ contains incongruent observers. 
In this case  ${S_p/\sim}$ has two, three or four classes ${\cal C}_p^{(1)},{\cal C}_p^{(2)},\dots,{\cal C}_p^{(N)}$, and no more than $4-N$ proper classes. Moreover, each one of this $N$ classes determines a unique positive real number $k_1,k_2,\dots,k_N$ such that if  $O, {\tilde O}\in{\cal C}_p^{(i)}$ then $O$ and ${\tilde O}$ are $k_i$-congruent.\\
\end{enumerate}

\noindent {\bf Summary.} As a brief summary of this section,  Theorem 3.1 means the following. Assume that we have a set of standard 
observers $S_p$ at $p$,  satisfying postulates $P1, P2$. Then each observer $O$ of $S_p$ is identifiable with a basis of tangent space $T_pM$, and we have one of the following three possibilities for the set of such bases:

\begin{enumerate}
\item There exist a unique $k$ such that any two $O, O' \in S_p$ are $k-$ congruent, that is, 
the corresponding transition matrix $A= M(Id,O\leftarrow{O'})$ belongs to the group 
$O^{(k)}(4,\mathbf{R})$. 

As we will see in more depth in the next section, one can assign a natural mathematical structure on $T_pM$ in this case. If, say, $k<0$, a Lorentzian scalar product $g_p$ to $T_pM$ is induced in the obvious way: simply, declare that the matrix of $g_p$ at any basis $O \in S_p$ is 
$I_3^{(k)}$. Analogously, when $k>0$ a Euclidean scalar product is induced, and in the limit cases $k= \omega (=\pm\infty), k=0$ a ``Leibnizian'' or ``anti-Leibnizian'' structure will be induced.

\item The second possibility is as the previous one but with the following difference: the value of $k$  is not unequivocally determined. In this case we have proven that necessarily all the matrices
 $A= M(Id,O\leftarrow{O'}),  \; O, O' \in S_p$ belong to the group 
$\{\pm 1 \} \times O(3,\mathbf{R})= \cap_k O^{(k)}(4,\mathbf{R})$ (that is: the observers at $S_p$ are $k-$congruent for any $k$). 

Thus, if a value of $k$ is chosen, the corresponding mathematical structure in previous case can be assigned (in fact, we also have a product structure in $T_pM$).

\item The third possibility is the existence of (at least) two {\rm incongruent} observers $O, O' \in S_p$, that is: the transition matrix $A= M(Id,O\leftarrow{O'})$ does not belong to the group 
$O^{(k)}(4,\mathbf{R})$ for any $k$. This case is scarcely representative (see Section 5) but possible (as we will see in Example 4.1). However, it can be controlled completely:

We have shown that there are at most four observers $O^{(i)}, i=1,2,3,4$ 
such that each two of them are incongruent. What is more, a positive number 
$k_i$ (say $0<k_1<k_2<k_3<k_4<\infty$) can be canonically associated to each one of these incongruent observers.

Therefore, if we consider a fifth observer $O^{(5)} \in S_p$ then it will be $k$--congruent to one of the four observers, for example $O^{(1)}$. We have then two possibilities for  the matrix $A= M(Id, O^{(1)} \leftarrow{O^{(5)} })$: (i) $A$ belongs to $\{\pm 1\} \times O(3, \mathbf{R})$ or (ii) $A$ does not belong to this group but it belongs to $O^{(k)}(4,\mathbf{R})$ for $k=k_1$. At any case, $A$ is $k_1 -$congruent. 

Thus, each one of the (at most) four incongruent observers $O^{(i)}$ determines a class of $k_i-$congruent observers ${\cal C}_p^{i}$. And an Euclidean product can be assigned canonically to each one of these classes.  Summing up, {\rm if there exist incongruent observers at $p$ then two, three or four Euclidean products can be canonically associated to $T_pM$.}

We go even further and distinguish two kinds of classes ${\cal C}_p^i$: the {\rm improper} classes (the matrix between two observers in the class always belongs to $\{\pm 1 \} \times O(3,\mathbf{R})$) and the {\em proper} classes (at least one of these matrices $A$ does not belong to $\{\pm 1 \} \times O(3,\mathbf{R})$ ---but necessarily $A \in O^{(k)}(4,\mathbf{R})$ with $k=k_i$). And the number of proper classes is also bounded: if there are four classes, none of them is proper; if there are three, at most one is proper; if there are two, the two classes may be proper. 

\end{enumerate}

\noindent {\bf Remark 3.1.} Theorem 3.1 has been proven for $S_p$ and, of course, it holds for $S^*_p$ too. If $S_p$ is not residual, Theorem 3.1 suggests that  
the transformations among basis in $S^*_p$ (or, more properly, in  $S^{*i}_p$, Definition 2.1)
have a  group structure
that is, in a natural way,
either one of the groups $O^{(k)}(4,\mathbf{R}), k\in S^1$ or the group $\{\pm1\}\times O(3,\mathbf{R})$ acts freely and transitively on  $S^{*i}_p$. This is always true except in the following case. Assume that  $S_p$ is equal to a unique proper class with  a certain associated $k$ (thus 
$S_p$ lies in the regular case). Recall that  $S^*_p$ was defined as the intersection of all the maximal sets of compatible observers containing $S_p$. Of course, if $k$ is not a positive real number clearly $S^*_p$ is the set of all the  observers which are 
$k-$congruent to those in $S_p$, and $O^{(k)}(4,\mathbf{R})$ acts on $S^*_p$. 
But if $k>0$ the following possibility also may appear: among the maximal sets of compatible observers containing $S_p$, there exist a set which contains incongruent observers (and, so, this set does not contain all the observers which are $k-$congruent to those in $S_p$).\\

\noindent {\bf Definition 3.1.} A point $p\in M$ will be called {\em residual}  if
neither one of the groups $O^{(k)}(4,\mathbf{R}), k\in S^1$ nor the group $\{\pm1\}\times O(3,\mathbf{R})$ acts freely and transitively on~$S^{*i}_p$.\\

\noindent Of course, 
if $p$ is not residual, then $S^{*i}_p$ is a embedded submanifold of $L_pM$.

\section{Mathematical models of spacetimes}
Theorem {\rm 3.1} combined with {\rm P3} permits the assignation of geometrical structures on the whole manifold, naturally associated to standard observers. First of all, it should be noted that if no residual points exist, P3 prohibits the simultaneous existence of two points $p,q\in M$, the first one in the regular case and the second in the degenerate case. In fact, otherwise $S^*_p$ would be diffeomorphic to  six-dimensional $O^{(k)}(4,\mathbf{R})$ for some $k \in S^1$, and $S^*_q$ would be diffeomorphic to three-dimensional $\{\pm 1\}\times O(3,\mathbf{R})$; but these dimensions must vary continuously if $S^* (\equiv S^*_U, U=M)$ is a submanifold of $LM$. This justifies the reason for dividing our study into the following three cases. 

\subsection{All the points are regular} 

Taking the value of $k(p)$ at each point one obtains a differentiable function $k: M\rightarrow S^1$ or {\em scale function}. We consider the open subsets $U_+ = k^{-1}(0,\infty), U_-=k^{-1}(-\infty , 0)$ and the closed ones $U_0=k^{-1}(0), U_\omega = k^{-1}(\omega)$.

At each point $p$ of $U_-$, the $k(p)$-congruence permits the construction of a Lorentzian scalar product $g_p$ by means of:
\begin{equation}
g_p(e_0,e_0)=k(p),\,g_p(e_i,e_j)= \delta_{ij}\mbox{ and }g_p(e_0, e_i)=0,
 \forall i,j\in\{1,2,3\},\;\;\;\;\;
\end{equation}
where $B_p=(e_0,e_1,e_2,e_3)$ is any basis of $T_pM$ induced by a standard observer around $p$. On varying $p$ at $U_-$, one constructs a Lorentzian metric $g$ on this open subset. Given a transition matrix between two standard observers $A=M(Id, O \leftarrow {\tilde O})$, the triad  ${\mathbf a_v}/a_{00}$ is the relative velocity of ${\tilde O}$ with respect to $O$ (see also Section {\rm 5}, point 2). From this we deduce that $c(p):=\sqrt{-k(p)}$ is the supremum of the relative speeds between standard observers at $S^*_p$ or {\em speed of light}. Note that the standard observers at each point $p$ are {\rm not} exactly the orthonormal bases for $g_p$; to obtain these it is necessary to re-scale each temporal unit $e_0$ dividing it by $c(p)$. This, subsequently, could appear to be a not very natural distinction and one could possibly suppose that $c\equiv 1$. However, the possibility obtained for the function $k$ enables one to understand how the transition to other mathematical structures out of $U_-$ is carried out, and gives import to the question of whether the speed of light is constant at different points (see also Section {\rm 5} point 4;  compare with the ``absolute time function'' which appears usually in the references on signature changing metrics, e. g., Ref. \ref{KK}, Section 2).

In open subset $U_+$, the relations (\rm 13) allow the definition of a metric again, even though this metric is now Riemannian. Note that such possibility was totally predictable from {\rm P2} (independent symmetries between temporal and spatial coordinates do not exclude the possibility that symmetries between all the coordinates also exist ---which, at least when $k=1$, occurs at $U_+$).

At each point of closed subset $U_\omega,$ the first element $\Omega_p$ of the dual basis of each standard observer is determined unequivocally, except the sign, and in cotangent space one can assign a degenerated metric whose radical is spanned by $\Omega_p$. In consequence, at each space at rest $\langle\partial_{x^1}|_p,\partial_{x^2}|_p, \partial_{x^3}|_p\rangle
 = {\rm Kernel}\,(\Omega_p)$, a Euclidean scalar product $h_p$ is induced. 
In the interior  of $U_\omega$ (when non--empty) the spaces at rest form a Riemannian vector bundle (subbundle of the tangent bundle) $RM$. Furthermore, ignoring, as usual, the question that the sign of $\Omega_p$ is not well defined (that is to say, considering the problem locally, or taking an appropiate covering of two sheets, and fixing, at some point $p$, the sign of $\Omega_p$ or {\em temporal orientation}), the linear 1-forms $\Omega_p$ unite making an everywhere non-zero differentiable 1-form. It is natural to call this 1-form the {\em absolute clock}, and only when this 1-form is exact, $\Omega = dt$, will we call function $t$ {\em absolute time}, which is defined unequivocally, except an additive  constant (see also Section {\rm 5}, point 3). Both elements, Riemannian fiber bundle $RM$ and 1-form $\Omega$ are still well defined even on the boundary points of $U_\omega$ (at these points their differentiability is well defined thanks to the differentiability of $S^*$). Note that at each point of  $U_\omega$ the relative speeds are not bounded, therefore, its supremum is $c=\infty$. We call the pair $(\Omega,h)$ formed by an everywhere non-zero one form $\Omega$ and a Riemannian metric $h$ on the kernel subbundle of $\Omega$ {\em Leibnizian structure}, and we call $RM:=({\rm Kernel}\,(\Omega), h)$ the bundle of {\em spaces at rest}. As mentioned, the Leibnizian structures are natural generalizations of those which are usually assumed for Newtonian spacetimes.

The structures for the closed subset $U_0$ are dual to the previous ones. More specifically, the temporal units of all the standard observers at $p$ coincide (except the sign, once again) allowing the definition of a differentiable vector field $Z$ or {\em ether field}. Moreover, the spatial units permit the definition of a positive semidefinite metric $g$ on $M$, whose  radical is spanned by $Z$ (or, equivalently, a Riemannian metric $h^*$ on the kernel subbundle of $Z$ in cotangent space). Note that now all the relative velocities between standard observers are null  ($c=0$), that is, the ether field (which has the same coordinates for all standard observers) allows standard observers to measure {\em absolute} velocities (velocities with respect to ether field). We call the pair $(Z,g)$ formed by an everywhere non-zero vector field $Z$ on $M$ and a positive semidefinite metric $g$, whose radical is spanned by $Z$, an {\em anti-Leibnizian structure}.  Although this structure is completely analogous mathematically to the Leibnizian one, it may sound strange from a physics point of view. Nevertheless, it presents two interesting characteristics. First, the anti--Leibnizian structure is the more general and simple mathematical model which allows one to speak of absolute velocities, 
which, historically, appear to have been required by physics. The second, when one works with signature changing spacetimes one supposes that the transition between the Riemannian and Lorentzian parts is carried out using a metric on all the manifold which degenerates at a hypersurface, see for example, Ref. \ref{DEHM}, Ref. \ref{Em}, Ref. \ref{Ha} and Ref. \ref{KK}. But it is natural to think that the group of automorphisms of the mathematical structure assigned to each point of spacetime has to vary in a continuous way. Thus, one must add an additional mathematical structure on this hypersurface. This additional structure yields an anti-Leibnizian structure, naturally. On the other hand, recall that this transition could be done taking the 2-contravariant tensors associated to the Riemannian and Lorentzian metrics, and considering a Leibnizian structure on the hypersurface. 

\subsection{ All the points are degenerate } 

The standard observers now determine a natural division of tangent bundle into two subbundles $TM= \hbox{Tem}(M) \oplus \hbox{Sp}(M)$, the first one $\hbox{Tem}(M)$ is spanned by timelike units, the second one $\hbox{Sp}(M)$ is equal to the set of all spaces at rest. Moreover,  we also have: a Riemannian metric at each subbundle (even though in the first subbundle the metric can also be taken as negative definite, in order to induce a Lorentzian metric on all $TM$), an ether field $Z$, which spans $\hbox{Tem}(M)$, and an absolute clock 1-form $\Omega$, whose annihilator is $\hbox{Sp}(M)$; $Z$ and $\Omega$ are defined again up to sign, and satisfy  $|\Omega (Z)| = 1$. 
Of course, this mathematical model, with the additional hypothesis that the absolute clock $\Omega$ comes from an absolute time $t$, is (locally) the model of spacetime better adapted to the ideas of the physicists of XIX century, who tried to make compatible Newtonian kinematics and  Maxwell equations. 

\subsection{There are residual points } 

For residual points perhaps $S_p^{*i}$ is not a submanifold of $L_pM$. Then P3 would not be applicable, and no conclusion on differentiability could be obtained. Thus, strictly, if there is a residual point $p$ then: (i) there are canonically defined $N(p)\leq 4$ Euclidean products on $p$ and (ii) $p$ may be out the subset $U$ in P3, and, so, nothing can be said on differentiability for this $N(p)$ Euclidean products when $p$ varies.

Nevertheless, Proposition 3.1 yields a control so accurate  on residual points that we can wonder if it is not possible to control the differentiability even in this case too. 
In order to get it, a more accurate version  P3' of postulate P3 must be stated: some assumption on differentiability must also be imposed for the points in $M$ out the open subset $U$. 
Of course, we did not state this more accurate version  P3' in Section 2 for simplicity; recall that the existence of residual points is very speculative. Moreover, there are different possible choices of what ``differentiability'' must mean in residual points.
However, all this will be briefly analysed in the remainder of this section, in order to check that even this case can be controlled completely.

Roughly speaking, residual points can be studied as if they were regular, with the following  difference: even though there are standard observers with transformations not in $\{\pm 1\}\times O(3,\mathbf{R})$,  they are so few that a unique Riemannian metric cannot be fixed. Nevertheless, $N\leq 4$ Riemannian metrics $g_1, \dots, g_N$ are characterized, with $N$ associated positive functions $k_1, \dots , k_N$. So, if we want to impose  differentiability, it is natural 
to assume directly the differentiability of $g_1, \dots, g_N$ and $k_1, \dots, k_N$, rather than the differentiability of $S^{*i}_p$ with $p$. Nevertheless, there is more than one possibility to define differentiability in this case. One possibility would be the following:\\

\noindent {\bf P3'(I). }Consider the set $\bar S^{*i}=\cup_{p\in M} \bar S_p^{*i}$ 
where $\bar S_p^{*i}$ means: (i) if $p$ is not residual, the fundamental system of units at $p$, as above ($\bar S_p^{*i}= S_p^{*i}$), (ii) if $p$ is residual then $\bar S_p^{*i}$ is the set of all the  basis tangent to  $p$ which are orthonormal for some of the metrics $g_i$'s except for the squared norm of the first vector, which  is equal to $k_i$\footnote{Recall that $\bar S_p^{*i}$ is a embedded submanifold of $L_pM$ in both cases.}. Then, $\bar S^{*i}$ is a submanifold of $LM$. \\

\noindent Moreover, one could think that this definition is somewhat restrictive, because it excludes the possibility that two such metrics $g_i$'s can be equal on some subset $V$, splitting in the boundary of $V$. For example, a less restrictive assumption would be to consider directly differentiability of the $g_i$'s and $k_i$'s in the following sense: for each point $p$ where the $g_i$'s and $k_i$'s  are defined, there exist an open neighborhood $U$ of $p$ and $N(p)\leq 4$ differentiable metrics and functions on open subsets of $U$ such that at each $q\in U$ these metrics and functions (perhaps two or more of them equal at $q$) coincides with the $g_i$'s and $k_i$'s at $q$. This would yield an alternative possible extension, P3'(II) of postulate P3.

On the other hand, another alternative extension P3'(III) would be natural, but only in the following particular case. Assume that there is some reason to choose one of the metrics in all the residual points. This is not general, but it happens when there exist only one proper class at each residual point. Then P3'(III) states that only this priviledged metric is assumed to be differentiable.\\

\noindent {\bf Example 4.1.} (A) First, we are going to construct an example with residual points but only one proper class. Consider at each $p\in\mathbf{R}^4$ the observers $O^{(0)}$, determined by the usual basis at $p$ and $O^{(i)}, i=1,2,3$ determinated by the  matrices $A^{(i)}=M(Id,O^{(i)}\leftarrow O^{(0)})$:
\[
A^{(1)}= \left(
\begin{array}{c|c}
 0 & \matrix{1 & 0 & 0} \\ \hline
\matrix{1\cr
0\cr
0} & \matrix{0 & 0 & 0\cr
             0 & 1 & 0\cr
             0 & 0 & 1}
\end{array}\right), \quad 
A^{(2)}=\left(
\begin{array}{c|c}
 0 & \matrix{0 & 0 & 1} \\ \hline
\matrix{0\cr
0\cr
1} & \matrix{1 & 0 & 0\cr
             0 & 1 & 0\cr
             0 & 0 & 0}
\end{array}\right),
\]
\[ 
A^{(3)}=\left(
\begin{array}{c|c}
 0 & \matrix{0 & 1/\sqrt{k(p)} & 0} \\ \hline
\matrix{0\cr
1\cr
0} & \matrix{1 & \,\;\, & 0 & \,\;\, & 0\cr
             0 & \,\;\, & 0 & \,\;\, & 0\cr
             0 & \,\;\, & 0 & \,\;\, & 1}
\end{array}\right),
\]
where $k(p)>0$. For any $k(p)$ the four observers form a compatible set $S_p$ and, at the points where $k(p)=1$, all of them are orthonormal bases for the usual metric of  $\mathbf{R}^4$. Nevertheless, at the points where $k(p)\neq 1$ observer $O^{(3)}$ is not orthonormal (it has a bad normalization ``in its spatial coordinates''), and the unique proper class is ${\cal C}_p = \{ O^{(0)},O^{(1)},O^{(2)}\}$. Note that the value of $k$ assigned by Lemma {\rm 3.1} to the proper class is  1, and to $[O^{(3)}]$ is $k(p)$. Of course, in this example, the proper class always determines the usual Riemannian metric of $\mathbf{R}^4$, and  $[O^{(3)}]$ determines other Riemannian metric which, according to P3'(III), would be neglected. 

Recall that if $k(p)$ is a differentiable function equal to $1$ in some open subset $W$ but not constantly equal to 1, then there is a sudden change in  $S^{*i}_p$ at the boundary of $W$. Rigourosly, this case would be incompatible with  {\rm P3'(I)}, but can be admitted if we assume just {\rm P3'(II)}. Moreover, it can also be admitted assuming P3'(III) because the metric assigned to the proper class varies smoothly\footnote{The possibility shown in this example is quite general whenever a Riemannian metric is considered. In principle,  one can take orthonormal bases (with a normalization of the first vector not necessarily equal to 1) as standard observers, and this  seems to be the most natural choice. But it is also possible to choose  at some points not only orthonormal bases but also other observers, associated to non-orthonormal  bases, but satisfying {\rm P2}. This fact is a consequence of the additional symmetries among temporal and spatial coordinates in the Riemannian case. Thus, it has no  analog in the other cases, and it seems to have no interest neither from physics nor from mathematics point of view.}.

(B) Recall that if  either none or more than one (necessarily two) proper classes exist at a residual point $p$, then there is no a canonical consistent choice of the classes, and P3'(III) would not be applicable. Let us consider an example with no proper classes. Assume that  we take in the example above, $S_p=\{O^{(0)},O^{(3)}\}$. Now, it is as natural to choose the Riemannian metric for which  $O^{(0)}$ is an orthonormal basis, as to choose  the one such that so is $O^{(3)}$. There are two representative behaviors of $k(p)$ in this example. 

The first one, analogous to a case (A) above, is the following: $k(p)$ is differentiable and defined on all $\mathbf{R}^4$ and, at some proper subset $V \neq \mathbf{R}^4$, $k(p)$ is equal to 1. Then, we have  two options: (i) to admit this case, looking at the boundary points of $V$ as a place where the Riemannan metric splits, according to  P3'(II), or (ii) to reject it, applying strictly P3'(I). 

The second behavior is that function $k(p)$ (and observer $O^{(3)}$) is defined just on an open proper subset $W \subset M, W\neq M$ (being or not function $k(p)$ extendible out of $W$). This may happen even in such a way that all is compatible with {\rm P3'(I)}. Again, we have two options: 
(i) to admit this case; thus, there exists a second Riemannian metric on $W$, or (ii) to reject it, by choosing a more restrictive formalization of   P3', arguing that there is a too sudden change at the boundary of $W$.\\

\begin{flushright}
$\triangleleft$
\end{flushright}

\noindent In general, depending on the mathematical formalization of {\rm P3'} we have the following possibilities. Under the more restrictive one, if there exists a residual point $p \in M$ with $N \leq 4$ classes of equivalence, then {\rm all} the points are residual with $N$ classes, and these classes determine $N$ diferentiable functions $0<k_1< \cdots<k_N$ and $N$ Riemannian metrics $g_1,\dots,g_N$ on all $M$. 

Under the less restrictive formalization, the existence of residual points is compatible with the existence of regular points (but not with degenerate points). Thus, as in the regular case, we can consider the closed subsets $U_0,U_\omega$, and the open subsets  $U_-$, $U_+=M\backslash \{U_0\cup U_\omega\cup U_-\}$. The only differences with the regular case are that, now, for $U_+$: (a) we can have $N(p)\leq 4$ Euclidean products at each $p\in U_+$, with associated  $k_1,\dots,k_{N(p)}$, (b) the number of associated Riemannian metrics may vary from one point to another, and (c) it is possible that a Riemannian metric splits in a point. However, this happens differentiably and, eventually, the Riemannian metrics are connected to the Leibnizian and anti-Leibnizian structures in a differentiable natural way.

\section{Further discussion}

In principle, all conclusions deduced from Theorem {\rm 3.1} are equally valid and it is not the main aim of the current paper to select the best spacetime model. Indeed, only time and lenght measurements could be not useful for choosing the best one. However, it is interesting to point out the following aspects, which may suggest extra ``optional'' postulates. These postulates are independent of previous ones, and one can discuss about if they are essential to measure space and time. Notice that any of them excludes the existence of residual points.

\begin{enumerate}
\item 
Let  $O_1 ,O_2\in S^*_p$ be two standard observers and determine a new (instantaneous) observer $O_3$ by means of:
\[
M(Id,O_3\leftarrow O_2) = M(Id,O_2\leftarrow O_1).
\]
{\em Transitivity} holds if, necessarily $O_3 \in S^*_p$ (i. e.: if the role of standard observer $O_2$ is interchangeable to the role of $O_1$, in the class $S^*_p$ of equivalent observers  of $S_p$). It is clear that all the non-residual cases  satisfy transitivity. This feature is equivalent to saying that if $A$ is a transition matrix between two standard observers, $A^2$ must be necessarily another one. Thus, under this hypothesis, no residual case in Theorem {\rm 3.1} can hold. This yields a considerable simplification of the mathematical structure of $S^*_p$ and $S^*$. In summary: as a consequence of transitivity, $S_p$ does not lie in any residual case of Theorem 3.1; so, $S_p^*$ can be regarded as a differentiable manifold where either one of the groups $O^{(k)}(4,\mathbf{R}), k\in S^1$ or $ \{\pm 1\} \times O(3,\mathbf{R})$ acts free and transitively. Recall that a  group structure for transformations between observers has been assumed traditionally (for example, axiom (4) in Ref. \ref{St}). But, under our approach, this structure is a consequence of more basic fundamentals. 

In addition, it is not difficult to check that transitivity implies the equality $det^2A=1$. Nevertheless, this {\rm a priori} weaker condition ({\em conservation of the volume}) is {\em a posteriori} equivalent to transitivity.
\item 
Till now, the only difference between space and time is the fact that, crossed derivatives between temporal coordinates and spatial coordinates are independently calculated (see {\rm P2}). Nevertheless, it is reasonable to assume {\em temporality}; i. e.: the timelike coordinates of any two standard observers $O, {\tilde O} \in S^*_p $ satisfy $\partial_{\,t}\,{\tilde t}|_p \not= 0$ \footnote{The more restrictive assumption can be also accepted $\partial_{\,t}\,{\tilde t}|_p > 0$, or {\em temporal orientation}. Under this assumption, the absolute clock 1-form $\Omega$ and the ether field $Z$ are unequivocally defined (not only up to a sign).}. This feature also admits the following interpretation. For each standard observer $O$, taking coordinates $(U,\Phi)$, the differentiable curve, $s\mapsto\Phi^{-1}(s,0,0,0)$ is the ``observer's trajectory'' into the spacetime, which is parameterized by its own time. Given a second standard observer ${\tilde O}$, the {\em relative trajectory} of ${\tilde O}$ measured by $O$ is the reparameterization of the curve ${\tilde s} \mapsto{\tilde\Phi}^{-1}({\tilde s},0,0,0)$ using the temporal coordinate of $O$. This reparameterization is always differentiable (at least in a neighborhood of $0 = \Phi^{-1}(p)$) if and only if, temporality holds.

It is interesting to emphasize that if this feature holds, then $a_{00}\not= 0$ for every $M(Id,O\leftarrow{\tilde O})$, $O,{\tilde O}\in S^*_p$. Thus, not only the residual cases but also the Riemannian case are excluded. Consequently, temporality is more restrictive than transitivity. Nevertheless, the variation of $c\in[0,\infty]$ is allowed and so, the corresponding mathematical structures (at each point of spacetime) are also permitted. Additionally, for each matrix $M(Id,O\leftarrow{\tilde O})$, the quotient ${\mathbf a_v}/a_{00}$ corresponds to the velocity of ${\tilde O}$ observed from $O$ (the velocity of the relative trajectory). Thus, the previously commented  characterization of $c$, as {\em supremum of the relative velocities for standard observers}, becomes apparent. The accepted name {\em speed of light} should be justified latter in the context of electrodynamics theory.
\item
When temporality is assumed, one wonders about {\em synchronizability} in the following sense. In Lorentzian Geometry, a timelike unit vector field $Z$ is {\em locally proper time synchronizable} (LPTS) if, for any point $p$, there exist a function $t: U \rightarrow \mathbf{R}$ on some neighborhood $U$ of $p$ whose gradient is $Z$. Or, equally, when the 1-form $\omega$ metrically equivalent to $Z$ is closed, $d\omega =0$. When only $\omega \wedge d\omega = 0$ holds, $Z$ is called locally synchronizable (LS); this is equivalent to the fact that the  orthogonal distribution to $Z$ is involutive. It is also equivalent to the local equality $\omega = -h dt$ for functions $h, t$ ($h$ positive) on $U$ (see, e. g., Ref. \ref{SW}, Section 2.3). If $Z$ is LPTS, or even just LS, the tangent spaces to the level hypersurfaces of $t$ (orthogonal distribution to $Z$) are the spaces at rest of the standard observers with the temporal axis spanned by $Z$. It is not difficult to prove that any Lorentzian manifold is LPTS, in the sense that for any point into the Lorentzian manifold, there is a neighborhood which admits a LPTS timelike unit vector field (of course, the neighborhood also admits timelike unit vectors which are not LPTS). Nevertheless, in a Leibnizian manifold, standard observers determine directly just one 1-form $\Omega$ (up to a sign). Thus, a unique distribution of spaces at rest, is also determined. So, a Leibnizian spacetime is said {\em LPTS} (resp. {\rm LS}) when $d\Omega = 0$ (resp. $\Omega \wedge d\Omega = 0$). 

Into the commonly accepted definition of ``Newtonian spacetime'' one assumes not only the condition LPTS, but also that $\Omega$ is exact (the  differential of a globally defined ``absolute time''). However, if a Leibnizian structure must be the ``limit'' of a Lorentzian one, the former is LPTS just for some cases.
\item
Nothing has been noted about how standard observers at two different points can be related --``gravitation'' has not been defined yet. Nevertheless, for the Lorentzian and Riemannian cases, covariantly constant metrics allow the parallel transport of standard observers. When the scale function, $k$, is not constant, these connections do not transport standard observers into standard observers (timelike units must be re-scaled). {\em Normalization} holds when the function $k$ is constant on all the manifold. In this case, changes in the mathematical structures are forbidden, and covariantly constant metrics are the prospective spacetime gravitation theories.

On the other hand, note that there is not a unique connection associated to a Leibnizian structure (resp. anti-Leibnizian structure). An obvious requirement for such a connection is to be {\em Galilean}, that is, it must transport standards observers into standard observers. Regarding $S^*$ as a principle fiber bundle with associated structural group $O^{(\omega)}(4,\mathbf{R})$ (resp. $O^{(0)}(4,\mathbf{R})$), a connection on $M$ is Galilean if and only if it comes from a connection into this bundle. One can check that a Leibnizian structure supports a {\em symmetric} (in the ordinary sense) Galilean connection if and only if it is LPTS. But in this case the symmetric connection is not unique, which contrasts with the Lorentzian or Riemannian cases. In the usual definition for Newtonian spacetime, the existence of a more restrictive connection is assumed, where a considerable part of the curvature tensor vanishes (see, e. g., Ref. \ref{MTW}, Box {\rm 12.4}). Such a connection not necessarily exist on a general Leibnizian spacetime, even in the LPTS case. The discussion of the possible connections associated to spacetimes will be tackled in a future paper, Ref \ref{BS}.
\end{enumerate}

\section{Conclusion}

We have introduced a minimal postulational basis for any non--quantum theory of spacetime. This is previous and simpler than any other approach (in particular EPS), and applicable to Einstein as well as Newton theories. It is also compatible with less standard approaches, as those in Ref. \ref{EM} and Ref. \ref{ZTE}. 

Our basic postulate, P2, ensures just the possibility that, at the end, standard instruments will be able to be constructed. Of course, once P2 is admitted, one has the experimental problem of finding the postulated standard observers. Now, it makes sense to wonder, for example, if such observers must take into account necessarily some types of forces (gravity, electromagnetism). At this level, axioms as those in EPS are useful. However, it seems that they can be simplified or better understood now. In fact, EPS is obviously compatible only with a Lorentzian structure (and not with either a  Leibnizian, or anti-Leibnizian or Riemannian structure). Thus, at least the EPS condition of compatibility with the metric, Ref. \ref{Eh}, formula (2.21), becomes now completely natural\footnote{In Ref. \ref{Eh}, p. 36 last line, p. 37, first one the condition of compatibility is said to be ``not satisfactory; it is an extraneous element of the theory''.}: {\em the unique possibility compatible with the other EPS axioms, among our final four structures, is   a Lorentzian metric.}

Moreover, it is remarkable that we obtain only a few structures, and the possible transitions between them. For example, the possibility of  ``variation of speed of light'' has been studied recently by some authors (see, Ref. \ref{St}, and references therein), and our postulates can be compared with the axioms or prescriptions in these references (for example, Ref. \ref{St}, Section III; recall that ours apply directly to the fundamental step of ``possibility of agreement between observers'').

Recall also that, in the last decade, considerable interest has been focused on the signature-changing metrics proposed by Hartle and Hawking, Ref. \ref{HH}, see for example, Ref. \ref{DEHM}, Ref. \ref{Em}, Ref. \ref{Ha} and Ref. \ref{KK}. Although Hartle and Hawking's ``no boundary proposal'' is related to quantum problems, the present paper try to open a new and more classical viewpoint. Notice also that we obtain two types of degenerations of the metric: the ``usual'' degeneration (studied in previous references) with a kernel spanned by a non-zero vector, and the ``dual'' degeneration, formulated in dual space. Furthermore, new structures are obtained over those degenerated regions, in which the authomorphism group dimension is constant over the manifold.

Finally, for the quantum case, the first question to solve would be: are the coordinate systems for the non--quantum spacetime theories also valid for quantum theories? Of course, there are possible negative answers, even though they are usually rather speculative. But notice that spacetime is a rather classical scenario in current Quantum Mechanics, and even  the actual origin of the quantum non--locality is founded in local processes, Ref. \ref{BM}. This seems to point out a positive answer. In this case, our models of spacetime would be the unique possible ones for any physical theory of spacetime.

\section{Acknowledgments}
We thank to Prof. Alfonso Romero and Prof. Antonio Fern\'andez-Barbero for kind comments on this work and constant encouragement. Suggestions and critiques by the referees have improved the paper, and they are acknowledged. This work was partially supported by MEC grant PB97--0784--C03--01.

\section{References}
\begin{enumerate}
\item\label{BS} A.N. Bernal and M. S\'anchez, (work in progress).

\item\label{BM} H. von Borzeszkowski and Michael B. Mensky, 
{\it Phys. Lett.} {\bf A269}, (2000) 204.

\item\label{DEHM} T. Dray, G. Ellis, C. Hellaby, and C.A. Manogue, {\it Gen. Relat. Gravit.} {\bf 29}  (1997) 591.

\item\label{Eh} J. Ehlers: ``Survey of General Relativity'' in {\it Relativity, Astrophysics and Cosmology} (Ed. W. Israel, D. Reidel Pub. Co.,  Dordrecht (Holland)/ Boston (USA), 1973).

\item\label{EPS} J. Ehlers, F.A.E. Pirani, and A. Schild, {\it General Relativity} (Ed. L. O'Raife\-artaigh, Oxford, 1972) 63.

\item\label{EM} G.F.R. Ellis and D.R. Matravers, {\it Gen. Relat. Gravit.} {\bf 27}  (1995) 777.

\item\label{Em} F. Embacher, {\it Phys. Rev. D} {\bf 51} (1995) 6764.

\item\label{Ha} S.A. Hayward, {\it Class. Quant. Grav.} {\bf 9} (1992) 1851.

\item\label{HH} J.B. Hartle and S.W. Hawking, {\it Phys. Rev. D} {\bf 28} (1983) 2960.

\item\label{KK} M. Kossowski and M. Kriele, {\it Class. Quant. Grav.} {\bf 10} (1993) 1157.
     
\item\label{Lo} A.A. Logunov, {\it Relativistic theory of gravity} (Nova Science Pub. Inc. Huntintong, New York, 1998).

\item\label{M} K.B. Marathe, {\it J. Diff. Geom.} {\bf 7} (1972) 571.

\item\label{MTW} C.W. Misner, K.S. Thorne, and J.A. Wheeler, {\it Gravitation,} (Freeman, San Francisco, 1973).

\item\label{SW} R.K. Sachs and H.H. Wu, {\it General Relativity for Mathematicians} (Springer-Verlag, N.Y., 1977).

\item\label{Sp} M. Spivak, {\it A comprehensive introduction to Differential Geometry}, vol. 2,   (Publish or Perish Inc., Boston, 1970).

\item\label{St} S.S. Stepanov, {\it Phys. Rev. D} {\bf 62} (2000) 0235071. 

\item\label{Wa} F.W. Warner, {\it Foundations of differentiable manifolds and Lie groups}, (Scott, Foresman and Co., Glenview, Illinois, 1971).

\item\label{ZTE} R. Zalaletdinov, R. Tavakol, and G.F.R. Ellis, {\it Gen. Relat. Gravit.} {\bf 28} (1996) 1251.

\end{enumerate}
\end{document}